\documentstyle [12pt]{article}
\begin{document}
\baselineskip = 21 pt
\thispagestyle{empty}
\title{
\vspace*{-1.5cm}
\begin{flushright}
\begin{tabular}{c c}
& {\normalsize MPI-Ph/93-25}\\
& {\normalsize April 1993}
\end{tabular}
\end{flushright}
\vspace{1.5cm}
Unification of Couplings and the Dynamical
Breakdown of the Electroweak Symmetry\thanks{To appear
in the Proceedings of
{\it Properties of SUSY Particles},
Erice, Italy, September 28-October 4, 1992.}
}
\author{ C. E. M. Wagner$^*$
 ~\\
 ~\\
Max Planck Institut f\"ur Physik \\
F\"ohringer Ring 6, D-8000
M\"unchen 40, Germany
{}~\\
{}~\\
{}~\\
{}~\\}
\date{
\begin{abstract}
I discuss the properties of the minimal supersymmetric extension
of the standard model, in the case in which there is a dynamical
breakdown of the electroweak symmetry induced by the formation of
condensates of the third generation of quarks and their
supersymmetric partners.  The top quark and Higgs
mass predictions derived within
this scheme are  essentially equivalent to those ones
obtained from
the requirement of bottom-tau Yukawa coupling
unification in a supersymmetric
grand unified scenario, if
the compositeness scale is identified with the grand unification
scale. I give an explanation of this interesting result, for
which the relevance of the infrared quasi fixed point on the
top quark Yukawa coupling is emphasized.
\end{abstract}}
%  *$ Address after October 1st,1993:
% Theory Division, CERN, 1211 Geneva 23, Switzerland.\\
\maketitle

\newpage
{\bf \hspace{-0.3in}{ 1. The
Top Condensate Model}}

\par
The increasing lower bound on the top mass has open the window for
a top quark heavy enough to induce the formation of
a condensate, which catalyzes the electroweak
symmetry breakdown at low energies$^{1-3}$.
In fact, in analogy to what happens in the
Nambu Jona Lasinio (NJL) model,
a strong Yukawa coupling could be the signature of a dynamical
mechanism  for the electroweak symmetry breaking
which relies only on the observed quark and leptons of
the standard model, and in which the Higgs field appears as
a $t - \bar{t}$ bound state. The basic mechanism for the
physical realization of this idea  was first  proposed by Nambu$^1$,
by making an analogy between the spontaneous
breakdown of the electroweak symmetry in the Standard Model and
the  BCS  mechanism in condensed matter theories.
Several authors analysed the physical consequenses of such a
scenario $^{2,3}$,
 and a detailed field theoretical analysis was first done
by Bardeen, Hill and Lindner$^{4}$. They started with
a gauged, $SU(3)_C \times SU(2)_L \times U(1)_Y$ invariant,
NJL model,
\begin{equation}
{\cal{L}} = {\cal{L}}_K^{\psi} + {\cal{L}}_{YM} +
G \left(\bar{\psi}_L^c t_R^c \right) \left( \bar{t}_R^d \psi_L^d
\right)
\end{equation}
where ${\cal{L}}_K^{\psi}$ and ${\cal{L}}_{YM}$ are the kinetic
terms for the fermion and Yang Mills fields respectively,
$\psi_L^T = (t \;\; b)_L$, $t$ and $b$ are the bottom and top
quark fermion fields and the indices $c$ and $d$ indicate
a sum over color degrees of freedom.  In this first simplified
formulation only the top quark acquires mass. The masses for the
other fermion fields, however, may be generated by introducing
the corresponding Yukawa couplings between the fermions and
the scalar composite field.
If $G > 0$ the interactions are attractive, and for $G > G_c$,
the local chiral symmetry of the theory is broken through
a top condensate, $< \bar{t} t > \neq 0$.  In the scaling region,
a composite scalar doublet,
\begin{equation}
H \; = \; G \; \bar{t}_R \psi_L
\end{equation}
appears in the spectrum of the theory. The quantum numbers of
these composite fields are exactly equal to those ones of the
elementary Higgs field in the standard model, and hence,
for $G > G_c$
the  $SU(2)_L \times U(1)_Y$ local symmetry is broken to
$U(1)_{em}$.  Three massless Goldstone bosons, associated with
the breakdown of the gauge symmetry are induced,
giving masses to the electroweak gauge bosons through the
usual Higgs mechanism. In addition, a physical, electrically
neutral scalar field appears in the  spectrum of the
theory. In general, the low energy spectrum is completely
equivalent to the Standard Model one, although the reduction in
the number of free parameters of the theory increases its
predictability.  In fact, as I shall discuss below,
for a given effective cutoff scale
$\Lambda$, sharp predictions for the scalar Higgs and top
masses can be derived within this context. \\

{\it \hspace{-0.3in}{ 1.1
Large $N_C$ Analysis.}}

 The dynamical properties of the gauged NJL model
can only be explored by using nonperturbative methods. A
systemathical analysis can be done, for example, by solving the
self consistent Schwinger Dyson equations of the theory in the
large $N_C$ approximation, where $N_C$ is the number of
 colors$^{4}$.
The critical four Fermi coupling may be estimated
by solving the self consistent equation  for the top quark
mass.
I shall first study the model in the so called bubble approximation,
that is the large $N_C$ limit, for vanishing $SU(3)_C$ gauge coupling
value. The dynamical
effects due to the inclusion of the $SU(3)_C$ interactions
will be discussed below. In the bubble approximation, the
self consistent equation for the top quark mass reads
\begin{equation}
M_t = \frac{ G N_C}{8 \pi^2} \left( \Lambda^2 -
M_t^2 \log \left(\frac{ \Lambda^2}{M_t^2} \right)\right)M_t.
\end{equation}
Hence, for a nontrivial solution of the gap
equation, $M_t \neq 0$, the top quark mass is given by
\begin{equation}
M_t^2 \log \left( \frac{\Lambda^2}{M_t^2} \right) =
\Lambda^2 - \frac{8 \pi^2}{N_C G}
\end{equation}
Observe that, since the left hand side is positive a nontrivial
solution only exists if $G > G_c$, with $G_c = 8 \pi^2/N_C
\Lambda^2$.
Since the logarithmic factor  is only a slowly varying function
of $M_t$, the natural scale for the fermion mass in the broken
phase would be just the
cutoff scale. A large hierarchy  between the cutoff
scale and the fermion mass scale requires a very precise fine tuning
of the four Fermi coupling to its critical value. This is nothing
but the usual fine tuning problem of the standard model.

 The relation between this model and the standard Higgs Yukawa
model becomes apparent if I rewrite the Lagrangian density in an
equivalent form, by  introducing  an auxiliary scalar
doublet $H$
\begin{equation}
{\cal{L}} = {\cal{L}}_K^{\psi} + {\cal{L}}_{YM}
+ \bar{\psi}_L^b t_R^b H + H^{\dagger} \bar{t}_R^b \psi_L^b
- M_0^2 H^{\dagger} H.
\end{equation}
In the above,
$M_0^2 = 1/G$ and $H$ can be eliminated through a Gaussian
integration, or equivalently by its replacement through
 its equation of motion
$H = G \; \bar{t}_R \psi_L$. At this level, the scalar field $H$ is
a static field, with no independent dynamics.
 The physical picture changes, however,
once the quantum fluctuations of the fermion fields are taken
into account.  In the bubble approximation, for example, the
scalar fields propagate through fermion
\lq\lq bubbles \rq\rq.
The propagator of the scalar
field $H$, $D^{-1}(p)$,  may be obtained
by computing the bubble function with external momentum $p$
through the relation
\begin{equation}
D^{-1}(p) = \frac{1}{G} + B(p)
\end{equation}
where $B(p)$ is the bubble function. A nonvanishing kinetic term
for the unrenormalized scalar field $H$ is induced, together with
a correction to its physical mass. The function $B(p)$ is
quadratically divergent, but the quadratical divergences are
cancelled once the gap equation is taken into account. Observe
that the fermion mass is nothing but the vacuum expectation
value of the electrically neutral, CP even component of the
scalar field, $<H^0>$,
and hence, once the gap equation is fulfilled,
the quadratical divergences of the scalar propagator are
automatically cancelled.

The propagator of the neutral scalar field $H^0$ may be explicitly
computed, giving
\begin{equation}
D^{-1}(p) = \frac{N_C}{8 \pi^2} \left( p^2 - 4 M_t^2 \right)
\log\left( \frac{\Lambda^2}{M_t^2} \right) + \chi(p^2) ,
\end{equation}
where for $p^2 = {\cal{O}}(M_t^2)$ and $\Lambda \gg M_t$,
the function $\chi(p^2)$ is negligible. Hence, the neutral scalar
field propagator has a pole at
\begin{equation}
M_{H^0} = 2 \; M_t.
\end{equation}
I would like to emphasize that the above
 prediction is obtained
in the context of the bubble approximation where important
effects, like the QCD corrections, are neglected.  Although
these corrections do not change the qualitative physical picture,
they have an important incidence on the quantitative relations
between physical couplings and masses.\\
% \newpage

{\it \hspace{-0.3in}{ 1.2 Effective Lagrangian Analysis.}}

 For $\Lambda \gg  M_t$, the values of the relevant quantities
are dominated by large logarithms, and all physical results
may be reproduced by doing an
effective field theory analysis$^{4}$.
I start with the Lagrangian density
\begin{equation}
{\cal{L}}(\Lambda) = {\cal{L}}_K^{\psi} + {\cal{L}}_{YM}
+ \bar{\psi}_L^b t_R^b H + H^{\dagger} \bar{t}_R^b \psi_L^b
- M_0^2 H^{\dagger} H,
\label{eq:LagLa}
\end{equation}
which characterizes the interactions at the large energy scale
$\Lambda$. The effective theory at the low energy
scale $\mu$ may be obtained by integrating out the short
distance fermion effects, which in this context is equivalent
to consider the quadratic and large logarithmic corrections
induced by the fermion loops. The effective low energy
Lagrangian reads,
\begin{eqnarray}
{\cal{L}}(\mu) = {\cal{L}}_K^{\psi} + {\cal{L}}_{YM}
+ \bar{\psi}_L^b t_R^b H + H^{\dagger} \bar{t}_R^b \psi_L^b
\nonumber\\
+ Z_H \left| {\cal{D}}_{\mu} H \right|^2 - \frac{ \lambda_0}{2}
\left( H^{\dagger} H \right)^2
- (M_0^2 + \Delta M^2) H^{\dagger} H,
\end{eqnarray}
where
\begin{eqnarray}
Z_H & = & \frac{N_C}{ (4 \pi)^2 } \log \left( \frac{\Lambda^2 }
{\mu^2} \right); \;\;\;\;\;\;\;\;\;\;\;\;\;\;\;\;\;\;\;
\lambda_0  =  2 Z_H,
\end{eqnarray}
while $\Delta M^2 \approx -1/G_c$.  The values of the wave
function renormalization constant and of the quartic couplings
are normalized so that the effective Lagrangian coincides with
Eq.( \ref{eq:LagLa})  at $\mu = \Lambda$. This leads to the
following boundary conditions,
\begin{equation}
Z_H ( \mu \rightarrow \Lambda ) = 0,\;\;\;\;\;\;\;\;
\lambda_0 (\mu \rightarrow \Lambda ) = 0,
\end{equation}
which are called the compositeness conditions.

The Lagrangian can be rewritten in a more conventional way by
normalizing the field $H$ so that it has a canonical kinetic
term, $H \rightarrow Z_H^{1/2} H$. In terms of the renormalized
field, it reads,
\begin{eqnarray}
{\cal{L}}(\mu) & = & {\cal{L}}_K^{\psi} + {\cal{L}}_{YM}
+ h_t \bar{\psi}_L^b t_R^b H
+ h_t H^{\dagger} \bar{t}_R^b \psi_L^b
\nonumber\\
& + &  \left| {\cal{D}}_{\mu} H \right|^2 - \frac{ \lambda}{2}
\left( H^{\dagger} H \right)^2
- m_H^2 H^{\dagger} H,
\end{eqnarray}
where the renormalized couplings $h_t = Z_H^{-1/2}$ and
$\lambda = Z_H^{-2} \lambda_0$. The compositeness conditions
imply the divergence of the renormalized couplings when
$\mu \rightarrow \Lambda$.

The physical Higgs and top quark masses
$M_{H_0}$ and $M_t$, (which in the absence
of gauge couplings coincide with the running ones, $m_{H_0}$
and $m_t$, respectively)
 are given by the on shell relations
\begin{equation}
m_t = h_t(m_t) \; v,\;\;\;\;\;\;\;\;\;\; m_{H^0}^2 =
2 \; \lambda(m_H^0) \; v^2
\end{equation}
where $v \simeq 175$ GeV,
is the vacuum expectation value of the renormalized
field. Since in the bubble approximation
the relation $\lambda(\mu)/(2 h_t^2(\mu)) = 1$
is fulfilled,  ignoring the small scale dependence, which
is of the order of other ignored higher order effects, the relation
$m_H^0 = 2 m_t$ is recovered. \\

{\it \hspace{-0.3in}{ 1.3 Improved Renormalization Group Analysis}}

The results of the last section can be improved by including
the electromagnetic and weak gauge
interactions, together with the dynamically generated scalar
effects. This can be done by including  nonleading order in
$1/N$ effects in the self consistent equations for the scalar Higgs
and top quark self energies.  When the compositeness scale is much
larger than the weak scale,
the value of the relevant coupling is
well determined by computing the leading logarithmic corrections. Hence,
the results of this approximation can be reproduced
by considering the full one loop renormalization group equations
of the standard model$^{4}$, while using the compositeness
conditions  discussed in the previous subsection,
 as an ultraviolet boundary condition at the compositeness
scale $\Lambda$,
\begin{eqnarray}
16 \pi^2 \frac{d h_t}{d t} & = &
\left(\left(N_C + \frac{3}{2} \right) h_t^2 -
(N_C^2 -1) g_3^2- \frac{9}{4} g_2^2 - \frac{17}{12} g_1^2
\right) h_t
\nonumber\\
16 \pi^2 \frac{d g_i}{ d t} & = & \beta_i g_i^3
\nonumber\\
16 \pi^2 \frac{d \lambda}{d t} & = & 12
\left( \lambda^2 + ( h_t^2- A)
\lambda + B - h_t^4 \right)
\label{Eq:renEq}
\end{eqnarray}
where $A = g_1^2/4 + 3 g_2^2/4$, $B = g_1^4/4 + g_1^2 g_2^2/8
+ 3 g_2^4/16$, $\beta_1 = - 41/6$, $\beta_2 = 19/6$ and $\beta_3
= 7$. Of course, the perturbative one loop renormalization group
equations may not be reliably used to determine the evolution
of the top quark
Yukawa coupling at scales close to the compositeness
scale $\Lambda$. However, the action of the infrared quasi-fixed
point makes the top quark mass predictions very insensitive
to the precise high value of the top quark Yukawa coupling at
the scale $\Lambda$.
In fact, for a compositeness scale of
the order of $\Lambda = 10^{10} - 10^{19}$ GeV, the
top quark Yukawa coupling
is strongly focussed to a small set of infrared values, with
corresponding running top quark masses of the order of $230$ GeV$^6$.
A slight variation, of less than $1\%$ ($2\%$)
of the top quark mass value is obtained by setting
 $h_t^2(\Lambda)/
4\pi = 1$, for a compositeness scale $\Lambda \geq 10^{16}$ GeV
($\Lambda \geq 10^{10}$ GeV).\\
% \newpage
{}~\\
\baselineskip=12pt
{\bf{Table 1.}}   Predictions for the top quark mass, $m_t$,
and the Higgs mass, $m_h$, in different approximations.\\
\baselineskip=21pt
\begin{center}
\begin{tabular}{|c|c|c|c|c|}
\hline
$\Lambda (GeV)\;\;\;\;\;\;\;\,$
&$10^{19}$  &$10^{15}$   &$10^{11}$  &$10^{7}$
\\  \hline
$m_t$ (GeV) Bubble Sum
 &144    &165    &200   &277
\\ \hline
$m_t$ (GeV) Planar QCD  &245    &262    &288   &349
\\ \hline
$m_t$ (GeV) Full RG Eq.   &218     &229   &248   &293
\\ \hline
$m_h$ (GeV) Full RG Eq.   &239    &256    &285   &354
\\ \hline
\end{tabular}
\end{center}
{}~\\
 The quartic coupling is also attracted to its infrared quasi
fixed point value, which as can be seen from Eq.(\ref{Eq:renEq}),
 gives a relation
between the top quark Yukawa coupling and the quartic coupling,
which translates into a mass ratio $m_{H^0}/m_t \approx 1.1$
The  numerical values for the top quark and Higgs masses
obtained in the different approximations and for different
values of the compositeness scale are
shown in Table 1$^{5}$.

The value for the top quark (Higgs) mass obtained by using the
full one loop renormalization group equations are stable under
variations of the compositeness scale $\Lambda$. It follows from
Table 1 that, starting from
$\Lambda \simeq 10^{19}$ GeV,
this mass value varies less than a 15 $\%$ ( 25 $\%$)
under a variation
of the compositeness scale of  eight orders of magnitude.  In
general, for  $\Lambda \leq 10^{19}$ GeV,
\begin{equation}
m_t \;> \; 210 GeV.
\end{equation}

Quite generally, for a given
effective cutoff scale $\Lambda$,
the triviality bound on the top quark
may be defined as the value
of $m_t$ which is obtained assuming that the top quark Yukawa
coupling becomes strong at scales of the order of $\Lambda$.
Since in the dynamical scheme under consideration the remormalized
coupling diverge at the compositeness scale,
the top quark mass obtained within the top condensate model
is consistent with the renormalization group trajectories
associated with the  triviality
bounds on this quantity, for an effective cutoff scale equal to the
compositeness scale.
The presence of the infrared quasi fixed point, makes the
value of this bound  very   insensitive to the exact large
value of the top quark
Yukawa coupling at the effective cutoff scale$^{5,6}$. Thus,
the values of the top quark mass derived above define the
triviality bounds on $m_t$ and
may be interpreted as the maximum allowed value of this quantity
in any theory in which no new physics appear up to scales of order
$\Lambda$. \\

{\bf \hspace{-0.3in}
{2. Supersymmetric Generalization and Unification of Couplings.}}

In spite of its beauty and simplicity,
there are two main problems in the standard formulation of the
top condensate model. The first one is that an unnatural fine tuning
of the four Fermi coupling is necessary in order to obtain a proper
physical spectrum. The second one, is the fact that, even for
a compositeness scale  of the order of the Planck scale
$\Lambda \simeq 10^{19}$ GeV, the
running top quark mass
turns out to be $m_t \simeq 220 GeV$, a value which could be
too large to be consistent with the experimental
constraints coming from the $\rho$ parameter measurement.
Supersymmetry provides a possible solution to these
problems. In a  supersymmetric extension of the
top condensate  model the quadratic
divergences disappear, and hence no fine tuning of the four
Fermi coupling constants is required$^7$.
In addition, the predicted top quark mass values are sensibly
lower than in the standard case$^8$.

Most interesting,
it has been also recently noticed that the values of the gauge
coupling constants measured at LEP are consistent with a
supersymmetric
grand unified scenario. Indeed, unification of couplings within
the Minimal Supersymmetric Standard Model (MSSM) may
be achieved if the grand unification scale is of the order of
$10^{16}$ GeV and the supersymmetric partners  masses, characterized
by a common mass scale $M_{SUSY}$, are of the order of the weak
scale$^{9,10}$.
 It is worth mentioning that the exact value of
the supersymmetric threshold scale necessary to achieve unification
of gauge coupling constants is strongly dependent on
the value of the strong gauge coupling,
$\alpha_3(M_Z)$ and the weak mixing angle, $\sin^2 \theta_W(M_Z)$.
Moreover,
as I shall   discuss below, when a
splitting of the supersymmetric partner masses is introduced,
the effective supersymmetric threshold scale
may be far away from the characteristic scale of the
supersymmetric mass spectrum$^{11-13}$.

In addition to the unification of gauge couplings, the unification
of the bottom quark and tau Yukawa couplings appears naturally
in many grand unified scenarios$^{13-17}$
At the one loop level
the bottom and tau Yukawa coupling renormalization group
equations depend only on the gauge couplings, which are
fixed by the unification conditions,  and the top quark
Yukawa coupling.
Hence, the requirement of Yukawa coupling unification,
together with the bottom quark and tau mass values, are
sufficient to  determine the top quark mass
as a function of $\tan\beta$, the ratio of vacuum expectation
values of the two Higgs doublets present in the theory.
This program was recently carried on by several authors.
One of the most interesting results is that, for
a running bottom quark mass $m_b(M_b) < 4.6$ GeV (which approximately
correspond to a physical bottom quark mass $M_b < 5.2$ GeV), the
top quark mass predictions are close to its quasi infrared fixed
point ones, associated with the
triviality bounds on this quantity$^{8,13,14}$.
Hence, for these values of the running bottom quark mass the
predictions of the grand unified scenario are remarkably
close to the ones
of the top condensate model with a compostiteness scale
$\Lambda \simeq M_{GUT}$. One of the purpose of this talk is
to explain the origin of this interesting coincidence. \\

{\it \hspace{-0.3in}{ 2.1 The Generalized Supersymmetric NJL Model}}

To describe the dynamics responsible for the top quark multiplet
condensation, I shall consider an
$SU(3)_C \times SU(2)_L \times U(1)_Y$
invariant  gauged supersymmetric Nambu-Jona-Lasinio
model$^{7,8}$,
with explicit soft supersymmetry breaking terms.
Written in terms of the two composite chiral Higgs superfields
$H_1$ and $H_2$,
the action of the gauged Nambu-Jona-Lasinio model
at the scale $\Lambda$ takes the form
%\newpage
\begin{eqnarray}
\Gamma_\Lambda  &=&
\int dV \left[ \bar Q e^{2V_Q} Q +T^C e^{-2V_T} \bar T^C
+ B^C e^{-2V_B} \bar B^C \right] (1-m_0^2 \theta^2 \bar\theta^2 )
\nonumber\\
&+& \!
\int \! dV  \left(\bar H_1 e^{2V_{H_1}} H_1
(1- M_H^2 \theta^2\bar\theta^2)\right)
 - \left[ \int \!dS \epsilon_{ij}
\left(\mu_0 H_1^i H_2^j (1 + B_0 \theta^2)
\right. \right.
\nonumber\\
&-&
\left. \left.
g_{T_0} H_2^j Q^i T^C ( 1 + A_0 \theta^2) \right) + h. c. \right],
\label{eq:action}
\end{eqnarray}
where $Q=\left({T \atop B}
\right)$ is the $SU(2)_L$ doublet of top and bottom
quark chiral superfields, $T^C~(B^C)$ is the $SU(2)_L$ singlet charge
conjugate top (bottom) quark  chiral multiplet,
$m_0$, $M_H^2$, $A_0$
and $B_0$ are soft supersymmetry breaking terms,
$dV = d^4x d\theta^2 d\bar{\theta}^2$ and
$dS = d^4x d\theta^2$. The quark
and Higgs multiplets
interact with the $SU(3)_C \times SU(2)_L \times U(1)_Y$
gauge fields via $V_i$. The usual superfield notation has been used.
At this level the superfield $H_2$
acts as a Lagrange multiplier, imposing  the
compositeness condition
\begin{equation}
  H_1 = \frac{g_{T_0}}{\mu_0} Q\;T^C.
\end{equation}
It is then straightforward to show that
the Nambu Jona Lasinio model
depends only on $\delta = A_0 - B_0$$^{8}$. Observe that the
supersymmetric generalization of the four Fermi interaction,
$\Gamma_{F}$,
is a $D$ term, which is automatically obtained when the
superfield $H_2$ is integrated out,
\begin{equation}
\Gamma_{F} = \frac{g_{T_0}^2}{\mu_0^2} \int dV
\bar{T^C}
\bar{Q}
e^{2V_{H_1}} Q T^C.
\end{equation}

In the presence of a condensate of top quark superfields, a
dynamical mass for the top quark  is generated. Its value
may be determined in a self consistent way by
using the Schwinger-Dyson equations in the bubble approximation.
In its simpler form, for $\delta=0$, the gap equation reads
\begin{equation}
G^{-1}= \frac{N_Cm_0^2}{ 16 \pi^2} \left[
\left(1+{ m_t^2  \over  m_0^2}\right)
\ln{\left(
\frac{\Lambda^4}{ \left(m_t^2 +m_0^2\right)^2 }\right) }-
\frac{2 m_t^2}{m_0^2}
\ln{\left(
\frac{\Lambda^2}{m_t^2} \right) } \right],
\end{equation}
where  $G= g_{T_0}^2/\mu_0^2$.  The usual quadratic
dependence on $\Lambda$, appearing in the standard top-condensate
model has been replaced by a mild quadratic dependence on the
soft supersymmetry breaking scale $m_0$ $^{7,8}$.
In general, the critical four Fermi coupling is of the order of the
largest soft supersymmetry breaking scale  appearing in Eq.(17).
Hence, as
we discussed above, for a soft supersymmetry
breaking scale of the order of
a few TeV,
no fine tuning of the four Fermi coupling is
necessary in this framework.

In general, for $\delta \neq 0$,
in the scaling region, in which the four Fermi coupling
constant is
close to its critical value, a gauge invariant kinetic
term for $H_2$ is induced at low energies.
Rescaling the field $H_2$, so that
it has a canonically normalized kinetic term, its low energy effective
action  is given by$^8$
\begin{eqnarray}
\Gamma_{H_2} &=& \int dV  \bar H_2 e^{2V_{H_2}} H_2 (1 + 2 m_0^2
\theta^2\bar\theta^2)
- \left[ \int dS \epsilon_{ij}
\left(\mu H_1^i H_2^j (1 + \delta \theta^2)
\right. \right.
\nonumber\\
&-&
\left. \left.
h_t H_2^j Q^i T^C  \right) + h.c. \right] + q.t.,
\label{eq:actz}
\end{eqnarray}
where I have defined the renormalized mass,
$\mu = \mu_0/\sqrt{Z_{H_2}}$,
and Yukawa coupling, $h_t = g_{T_0}/\sqrt{Z_{H_2}}$ with
the wave function renormalization constant $Z_{H_2} = \frac{g_{T_0}^2
N_C}{16 \pi^2} \ln{\frac{
\Lambda^2}{m_0^2}}$. In the above,
$q.t.$ represent the radiative corrections
to the quartic terms which will be analyzed below.
Since $\mu_0$ and $g_{T_0}$ have finite values, the above
renormalized couplings diverge at the scale $\Lambda$.
Observe that, although  the
cancellation of the supersymmetry breaking term $A(\mu)$
at all scales is only a property of the bubble
sum approximation, the relation $ A(\mu)|_{\mu \rightarrow \Lambda} = 0$
is a prediction of the model. It is also important to remark that
although
at high energy scales,
the mass parameter associated to $H_2$ is positive
due to the supersymmetric contribution proportional to
$\mu^2$, $m_2^2 = \mu^2 - 2 m_0^2$,
at low energies it tends to negative values,
inducing,  therefore, the breakdown
of the $SU(2)_L \times U(1)_Y$ symmetry.\\
{}~\\
{\bf\hspace{-0.3in}{ 3. Predictions of the SUSY Top Condensate Model
}}

\par  Instead of computing
gauge fields corrections and higher order in $1/N_C$ effects,
it proves convenient to
work with the full renormalization group equations of the
supersymmetric standard model$^{7,8}$.
The running top quark mass value is given by
$m_t = h_t(m_t) v_2$, where $v_i$ is the vacuum expectation
value of the scalar Higgs $H_i$.
The low energy value of the top quark Yukawa coupling
can be obtained by computing its renormalization group flow, using
the supersymmetric renormalization group equations for scales
$M_{SUSY} \leq \mu \leq \Lambda$ and those of the
Standard Model  with
one or two Higgs doublets, for $\mu \leq M_{SUSY}$, where $M_{SUSY}$
is the soft supersymmetry breaking scale$^8$.
At energy scales $\mu$ close to the
compositeness scale $\Lambda$,
the perturbative one loop renormalization group equations
can not be  used in a reliable way, to determine the evolution of the
top Yukawa coupling. However, the same as in the
standard case, the action of the
infrared quasi-fixed point yields the top quark mass
predictions quite insensitive to the precise high value of
the top quark Yukawa coupling at the scale $\Lambda$, or
more generally, to the inclusion of higher order
operators$^{8,18}$. In addition,
the running top quark mass is only slightly dependent on
the exact value of $M_{SUSY}$ and for a fixed compositeness scale,
it is  well approximated by the functional relation
$m_t \simeq M_T  \tan\beta/\sqrt{1+\tan^2\beta}$,
where $\tan\beta = v_2/v_1$. For
$\Lambda = 10^{16}$ GeV and a strong gauge coupling
$\alpha_3(M_Z) \simeq 0.12$, the value of the constant
$M_T$ is approximately given by
$M_T \simeq $ 195 GeV.

At energies below the  soft
supersymmetry breaking scale the  Higgs potential is given by the
general expression$^{8,20}$
\begin{eqnarray}
V_{eff} &=& m_1^2 H_1^{\dagger} H_1
+ m_2^2 H_2^{\dagger} H_2 - m_3^2 \left(
H_1^T i \tau_2 H_2 + h.c. \right)
+ \frac{\lambda_1}{2} \left( H_1^{\dagger} H_1 \right)^2
\nonumber \\
&+&\frac{\lambda_2}{2} \left( H_2^{\dagger} H_2 \right)^2
+ \lambda_3 \left( H_1^{\dagger} H_1 \right)
\left( H_2^{\dagger} H_2 \right)
+ \lambda_4 \left| H_2^{\dagger} i \tau_2 H_1^{*} \right|^2
\label{eq:corrpot}
\end{eqnarray}
where the radiative corrections induced by the top quark Yukawa
coupling$^{19}$  may be computed by solving
the corresponding renormalization group
equations for the quartic couplings.
At the supersymmetry breaking scale the quartic
couplings must fulfill the boundary conditions
\begin{eqnarray}
\lambda_{1}(M_{SUSY}) = \lambda_2(M_{SUSY})& = &
\frac{g_1^2 + g_2^2}{4},
\;\;\;\;\;\;\;\;\;\;\;\;\lambda_3(M_{SUSY}) = \frac{g_2^2-g_1^2}{4},
\nonumber\\
\lambda_4(M_{SUSY}) & = & -\frac{g_2^2}{2}.
\end{eqnarray}
If $m_A^0$, defined as
$(m_A^0)^2 = m_1^2 + m_2^2$,
is of the order of the weak scale, two light Higgs doublets
appear in the low energy spectrum.
There are two neutral
CP-even scalar states, one neutral  CP-odd state and
a charged state, whose masses may be obtained from the above
effective potential$^{8,20}$.

{}From the minimization of the
potential, a lower bound on $\tan\beta$
may be derived. It can be
shown that, under reasonable assumptions,
the characteristic values of the ratio of vacuum
expectation values $\tan\beta \geq 1$. Moreover, for a
characteristic soft supersymmetry breaking scale
 $M_{SUSY} = 1-10$ TeV and a compositeness
scale $\Lambda = 10^{10}-10^{16}$ GeV, the top quark mass
fulfills the condition $m_t > 140 GeV$$^8$.
For a given compositeness scale $\Lambda$ and
$M_{SUSY}$, the top quark mass is only a function
of $\tan\beta$,
while the Higgs spectrum  depends on $\tan\beta$ as well as
 on the
value of the mass parameter $m_A^0$.
Assuming $\tan\beta \geq 1$, $\alpha_3(M_Z) \simeq 0.12$,
the characteristic squark mass to be of the order
of 1 TeV and a compositeness scale $\Lambda = 10^{16}$ GeV the
upper bounds on the lightest Higgs mass within this model
 are $m_h \leq $ 65 GeV
if $\tan\beta \simeq 1$
($m_t \simeq 140$ GeV)
and $m_h \leq 135$ GeV if the ratio of
vacuum expectation values is
in the range $\tan\beta = 5 - 30$
($m_t \simeq 195$) GeV$^{8,20}$.
\\

{\bf\hspace{-0.3in}{
4. Unification of Couplings, $\alpha_3(M_Z)$ and $Y_t(M_{GUT})$}}

\par As we mentioned in the introduction, for given values of the
gauge couplings, the condition of unification
of bottom and tau Yukawa couplings allows to determine the value
of the top quark mass as a function of $\tan\beta$. The values
of the weak gauge couplings must fulfill very tight experimental
constraints. Indeed, working in the modified $\bar{MS}$
scheme$^{21}$,
the value of the fine
structure constant $\alpha^{-1}(M_Z) = 127.9$, while$^{12}$
\begin{equation}
\sin^2 \theta_W(M_Z) = 0.2324 \pm 0.006
\end{equation}
where the top quark mass value has been left free. The strong
gauge coupling value is not so precisely known and, a conservative
estimate for this quantity is$^{12}$
\begin{equation}
\alpha_3(M_Z) = 0.12 \pm 0.1,
\end{equation}
where the upper (lower) range of values are preferred by
LEP (deep inelastic scattering) data.
Since the top quark mass predictions coming from Yukawa
coupling unification depend on the value
of  $\alpha_3(M_Z)$,
a precise determination of the top quark mass can not be done
unless the value of $\alpha_3(M_Z)$ is  known. In the
following, for a given supersymmetric spectrum, and a given
value of the weak mixing angle,  the value of $\alpha_3(M_Z)$
will be determined by requiring the unification condition.
Within this context, the value of the strong gauge coupling
is given by$^{12,13}$
% \newpage
\begin{eqnarray}
\frac{1}{\alpha_3(M_Z)}  &=&
\frac{ \left(
b_1 - b_3 \right)}
{ \left(
b_1 - b_2 \right)} \left[
\frac{1}{\alpha_2(M_Z)}
 + \gamma_2  + \Delta_2 \right]
-
\frac{ \left(
b_2 - b_3 \right)}
{ \left(
b_1 - b_2 \right)}  \left[
\frac{1}{\alpha_1(M_Z)}
 + \gamma_1  + \Delta_1 \right]
\nonumber\\
& - & \gamma_3  - \Delta_3
+ \Delta^{Sthr}\left(\frac{1}{\alpha_3(M_Z)}\right) ,
\label{eq:pred3}
\end{eqnarray}
where
\begin{equation}
\Delta^{Sthr}  = \frac{19}{28 \pi}
\ln \left( \frac{T_{SUSY}}{M_Z}\right)
\end{equation}
is the contribution
to $1/\alpha_3(M_Z)$ due to the inclusion of the supersymmetric
threshold corrections at the one loop level, $\gamma_i$
includes the two loop corrections to the value of $1/\alpha_i(M_Z)$,
$\Delta_i$ are correction constants which allow to transform the
gauge couplings from the minimal $\bar{MS}$ scheme to the
dimensional reduction scheme, $\bar{DR}$, more appropriate for
supersymmetric theories, and $b_i$ are the supersymmetric beta
function coefficients associated to the gauge coupling $\alpha_i$.
The effective supersymmetric threshold scale
is defined as that one
which would produce the same threshold corrections to the value of
$\alpha_3(M_Z)$ in the case in which
all the supersymmetric particles were degenerate in mass.

In order to study the dependence of $T_{SUSY}$ on the
different sparticle mass scales of the theory,
we define
$m_{\tilde{q}}$,
$m_{\tilde{g}}$,
$m_{\tilde{l}}$,
$m_{\tilde{W}}$,
$m_{\tilde{H}}$ and $m_H$
as the characteristic masses of the
squarks, gluinos, sleptons, electroweak gauginos,
Higgsinos and  the heavy Higgs doublet,
respectively. Assuming different values for all these
mass scales, I derive an expression for
the effective supersymmetric threshold $T_{SUSY}$ which is given
by$^{13}$
\begin{equation}
T_{SUSY} =
m_{\tilde{H}}
\left( \frac{m_{\tilde{W}}
}{m_{\tilde{g}}}
\right)^{28/19}
\left[
\left( \frac{m_{\tilde{l}}}{m_{\tilde{q}}}
\right)^{3/19}
\left( \frac{m_H}{m_{\tilde{H}}}
\right)^{3/19}
\left( \frac{m_{\tilde{W}}}{m_{\tilde{H}}}
\right)^{4/19} \right] .
\label{eq:susym}
\end{equation}
The above relation holds  whenever
 all particles considered above have a mass
 $m_{\eta} > M_Z$. If, instead,
 any of the sparticles or the heavy Higgs boson
has a mass $m_{\eta} < M_Z$,
it  should be replaced
by $M_Z$ for the purpose of computing the
supersymmetric threshold
corrections to $1/ \alpha_3(M_Z)$.
In the following,  unless otherwise
specified, I shall assume that
all sparticles
and the heavy Higgs doublet acquire masses above $M_Z$.
{}From Eq.(\ref{eq:susym}), it follows that,
for fixed mass values of the uncolored sparticles,
that is sleptons, Higgsinos
and the weak gauginos, together with the heavy Higgs doublet,
the value of $T_{SUSY}$ decreases for larger mass
values of the colored
sparticles -
 squarks and
gluinos. Moreover,
$T_{SUSY}$ depends
strongly on the first two factors in Eq.(\ref{eq:susym}),
while it
is only slightly
dependent on the expression
 inside the squared brackets. This is most surprising, since it
implies that $T_{SUSY}$ has only a slight dependence on the
squark, slepton and heavy Higgs
masses and a very strong dependence on
the overall Higgsino mass, as well as on the ratio of
masses of the gauginos associated with the
electroweak and strong interactions. The mild dependence
of the supersymmetric threshold corrections on the squark and
slepton mass scales is in agreement with a similar observation
made in the context of the
 minimal supersymmetric SU(5) model$^{10}$.
In Table 2, I show the predictions for the strong gauge
coupling, for different values of $\sin^2\theta_W(M_Z)$ and
the supersymmetric threshold scale.\\
%\newpage
{}~\\
\baselineskip=12pt
{\bf{Table 2.}} Dependence of $\alpha_3(M_Z)$ on $\sin^2 \theta_W(M_Z)$
and $T_{SUSY}$,
in the framework of  gauge and bottom - tau
Yukawa coupling unification,
for $m_b(M_b) = 4.3$ GeV.\\
\baselineskip=21pt
\begin{center}
\begin{tabular}{|c|c|c|}
\hline \
$\sin^2 \theta_W(M_Z)$
&$\alpha_3(M_Z)$ for $T_{SUSY} = 1$ TeV
 &$\alpha_3(M_Z)$ for $T_{SUSY} = 100$ GeV
\\  \hline
$0.2335
$
 &0.111  &0.118
\\ \hline
$0.2324$
 &0.115  &0.122
\\ \hline
$0.2315$
 &0.118  &0.126
\\ \hline
\end{tabular}
\\
\end{center}

The issue  of unification of Yukawa couplings have been
recently analyzed in some detail$^{13-17}$.
It was shown that$^{13}$, for a given value
of the running bottom quark mass and the weak mixing angle,
the top quark Yukawa coupling at the grand unification scale
depends strongly only on the value of $\alpha_3(M_Z)$.
One of the most interesting results of this analysis
is that, for a running bottom quark mass
$m_b(M_b) = 4.3$ GeV, which approximately correspond to a
physical mass $M_b = 4.9$ GeV, the value of the top quark Yukawa
coupling at the grand unification scale must be much larger
than the gauge couplings$^{13}$. Moreover, the larger the value
of $\alpha_3(M_Z)$, the larger the value of $Y_t(M_{GUT})$
becomes. In Table 3, I present the predicted values of
the top quark Yukawa coupling at the grand unification scale
for $\sin^2\theta_W = 0.2324$, $m_b(M_b) = 4.3$ GeV and
different values of the effective supersymmetric threshold
scale $T_{SUSY}$.\\
{}~\\
\baselineskip=12pt
{\bf{Table 3.}} $Y_t(M_{GUT})$
predictions, as a function of $\alpha_3(M_Z)$
and $T_{SUSY}$, in the framework of gauge and bottom - tau
Yukawa coupling unification.\\
\baselineskip=21pt
\begin{center}
\begin{tabular}{|c|c|c|}
\hline \
$T_{SUSY}$[GeV] &$\alpha_3(M_Z)$   &$Y_t(M_{GUT})$
\\  \hline
$10^3$  &0.115   &0.3
\\ \hline
15   &0.127   &0.7
\\ \hline
7    &0.130   &1.0
\\ \hline
\end{tabular}
\end{center}
{}~\\
{}~\\
Observe that, for the above value of the bottom quark
mass the requirement of perturbative consistency of the top
quark Yukawa sector, $Y_t(M_{GUT}) \leq 1$ is sufficient to
constrain on the allowed value for the strong gauge
coupling. Indeed, in this case,
 the obtained upper bound  coincides
with that one coming from experimental limits on the
strong gauge coupling,
$\alpha_3(M_Z) \leq 0.13$.

The large value of the top quark Yukawa coupling necessary to
achieve unification of bottom and tau Yukawa couplings explains
why the top quark mass values predicted from the unification
condition quantitatively coincide with those ones  obtained
within the supersymmetric top condensate model. In fact, for
values of $Y_t(M_{GUT}) \geq 0.2$, and a grand unification
scale $M_{GUT} = {\cal{O}}(10^{16})$ GeV, the low energy values
of the top quark Yukawa coupling are strongly focussed to its
quasi infrared fixed point. I illustrate this behaviour in
Fig. 1, in which I plot the top quark mass as a function
of $\tan\beta$ for different values of the running bottom
quark mass in the range $m_b(M_b) = 4.1 - 4.6$ GeV (which
approximately correspond to the experimental allowed range for
the  physical bottom mass range
$M_b = 4.7 - 5.2$ GeV), and  characteristic values of
$sin^2 \theta_W(M_Z) = 0.2324$ and $\alpha_3(M_Z) = 0.122$.

Observe that, if in the minimal supersymmetric model
the physical top quark mass is below
$M_t \leq 160$ GeV, as may be inferred from precision measurement
analysis, then the running top quark mass $m_t \leq 152$ GeV.
As it may be seen from Fig. 1, such relatively low values of
the running top quark mass can only be obtained, in the framework
of gauge and bottom quark - tau Yukawa coupling unification, for
values of $\tan\beta$
close to one, or for very large values of $\tan\beta$$^{13}$.
Assuming moderate values of $\tan\beta$,
an upper limit on $\tan\beta  \leq 1.3$ may be obtained.
In addition, this bound implies strong constraints on
the Higgs sector of the theory. In fact, if the characteristic
squark mass is lower than or of the order of 1 TeV,
then the
lightest CP even mass will be
$m_h < 80$ GeV and it should be
observed at the LEP2 experiment$^{13,20}$.\\
{}~\\
\vspace*{11.5cm} \\
{}~\\
\baselineskip=12pt
{\bf{Fig.1.}} The predicted top quark mass as a function of
$\tan\beta$, assuming gauge and also bottom and
tau Yukawa coupling unification, for $m_b(M_b) = 4.6$ GeV
(dot-dashed line), $m_b(M_b) = 4.3$ GeV (solid line) and
$m_b(M_b) = 4.1$ GeV (dashed line).\\
{}~\\
\baselineskip=21pt
As illustrated in Table 4, the variation of
the running bottom quark mass in the range $m_b(M_b) =
4.1 - 4.7$ GeV, implies a large
variation of the top quark Yukawa coupling at the grand
unification scale. In fact, for the particular value of
$\alpha_3(M_Z)$ and $\sin^2 \theta_W(M_Z)$ considered in
Fig. 1 and Table 4, and for $m_b = 4.1$
GeV, the top quark mass predictions exactly coincide with
the ones of the SUSY top condensate models since
the top quark Yukawa coupling
$Y_t(M_{GUT}) = h_t^2(M_{GUT})/4 \pi$ acquires the maximum
allowed value consistent with a perturbative analysis of
the theory$^{13}$.\\
{}~\\
\baselineskip=12pt
{\bf{Table 4.}}  $Y_t(M_{GUT})$ predictions as a function of the
running bottom quark mass.\\
\baselineskip=21pt
\begin{center}
\begin{tabular}{|c|c|c|}
\hline \
$\sin^2 \theta_W(M_Z) = 0.2324$
&$\alpha_3(M_Z)$   &$Y_t(M_{GUT})$
\\  \hline
$m_b(M_b) = 4.6$ GeV
&0.122   &0.2
\\ \hline
$m_b(M_b) = 4.3$ GeV
&0.122  &0.5
\\ \hline
$m_b(M_b) = 4.1$ GeV
&0.122   &1.0
\\ \hline
\end{tabular}
\\
\end{center}
It is important to remark that, at
the grand unification scale, the characteristic value of the
top quark Yukawa coupling is five to ten times the gauge
coupling values.
This may only be avoided by chossing very large values of
$\tan\beta$, which, in the minimal $SU(5)$ model, are disfavoured
by proton decay constraints$^{22}$. As a matter of fact, the
existence of such  large values
of the top quark Yukawa coupling at the grand unification scale
provides a challenge for model builders. Most interesting, it
 might provide an
interrelation between the unification of couplings and the
minimal dynamical breaking of the electroweak symmetry within
the Minimal Supersymmetric Standard Model. \\
{}~\\
{\bf{Acknowledgements}}: Most of the results presented
above have been obtained
in collaboration with W. Bardeen, M. Carena, T. Clark, K. Sasaki
and S. Pokorski. \\


\newpage
\begin{thebibliography}{99}
\bibitem{Nambu} Y. Nambu,
{\it{Enrico Fermi Institute Preprint}} 89-08(1989).
\vspace*{-0.25cm}
\bibitem{Miransky}
V. Miransky, M. Tanabashi and K. Yamawaki, {\it{Phys. Lett.}}
{\bf{B221}} (1989)177.
\vspace*{-0.25cm}
\bibitem{Marciano} W.J. Marciano
{\it{Phys. Rev. Lett.}} {\bf{62}} (1989) 2793;
{\it{Phys. Rev.}} {\bf{D41}} (1990) 219.
\vspace*{-0.25cm}
\bibitem{BHL} W.A. Bardeen, C.T. Hill
and M. Lindner, {\it{Phys. Rev.}} {\bf{D41}} (1990) 1647.
\vspace*{-0.25cm}
\bibitem{Talks} C. Hill, {\it{Fermilab prep.}} FERMI-CONF-90/170, to
appear in the Proceedings of the Workshop on Dynamical Symmetry
Breaking, Nagoya (1990).\\
W. Bardeen, {\it{Fermilab prep.}} FERMILAB-CONF-90/269-T (1990).
\vspace*{-0.25cm}
\bibitem{Hill} C.T. Hill, {\it{Phys. Rev.}}
{\bf{D24}}(1981)691. \newline
C. Hill, C. N. Leung and S. Rao,
{\it{Nucl. Phys.}} {\bf{B262}}(1985) 517.
\vspace*{-0.25cm}
\bibitem{BCL}
W. A. Bardeen, T. E. Clark and S. T. Love,
{\it{Phys. Lett.}} {\bf{B 237}} (1990) 235.
\vspace*{-0.25cm}
\bibitem{Dyn} M. Carena, T.E. Clark, C.E.M. Wagner, W.A. Bardeen
and K. Sasaki, {\it{Nucl. Phys.}} {\bf{B369}} (1992) 33.
\vspace*{-0.25cm}
\bibitem{ABF} J. Ellis, S. Kelley and D.V. Nanopoulos,
{\it{Phys. Lett.}} {\bf{B260}} (1991) 131;\\
P. Langacker and M.X. Luo, {\it{Phys. Rev.}}
{\bf{D44}} (1991) 817;\\
F. Anselmo, L. Cifarelli, A. Peterman and
A. Zichichi, {\it{Nuovo Cimento}} {\bf{104A}} (1991) 1817;\\
U. Amaldi, W. de Boer and H. F\'urstenau, {\it{Phys.
Lett.}} {\bf{B260}} (1991) 447.
\vspace*{-0.25cm}
\bibitem{EKN} J. Ellis, S. Kelley and D. V. Nanopoulos,
{\it{Phys. Lett.}} {\bf{B249}} (1990) 441;
{\it{Phys. Lett.}} {\bf{B287}} (1992) 95;
{\it{Nucl. Phys.}} {\bf{B373}} (1992) 55;\\
F. Anselmo, L. Cifarelli, A. Peterman and A. Zichichi,
{\it{Nuovo Cimento}} {\bf{105A}} (1992) 581;\\
R. Barbieri and L. Hall, {\it{Phys. Rev. Lett.}}
{\bf{68}} (1992) 752;\\
J. Hisano, H. Murayama and T. Yanagida,
{\it{Phys. Rev. Lett.}} {\bf{69}} (1992) 1014.
\vspace*{-0.25cm}
\bibitem{RR} G. G. Ross and R. G. Roberts,
{\it{Nucl. Phys.}} {\bf{B377}}
(1992) 571.
\vspace*{-0.25cm}
\bibitem{Langacker} P. Langacker and N. Polonsky, {\it{U. of
Pennsylvania preprint}}, UPR-0513T, October 1992.
\vspace*{-0.25cm}
\bibitem{Nos} M. Carena, S. Pokorski and C.E.M. Wagner,
{\it{Max Planck Institut preprint}} MPI-Ph/93-10, February 1993.
\vspace*{-0.25cm}
\bibitem{Ramond} H. Arason, D. J. Casta\~no, B. Keszthelyi,
S. Mikaelian, E. J. Piard, P. Ramond and B. D. Wright,
{\it{Phys. Rev. Lett.}} {\bf{67}} (1991), 2933.
\vspace*{-0.25cm}
\bibitem{KLN} S. Kelley, J.L. Lopez and D.V. Nanopoulos,
{\it{Phys. Lett.}} {\bf{B278}} (1992) 140.
\vspace*{-0.25cm}
\bibitem{DHR} S. Dimopoulos, L. Hall and S. Raby,
{\it{Phys. Rev. Lett.}} {\bf{68}} (1992) 1984,
{\it{Phys. Rev.}} {\bf{D45}} (1992) 4192.
\vspace*{-0.25cm}
\bibitem{BB} V. Barger, M. S. Berger and P. Ohmann,
{\it{Phys. Rev.}} {\bf{D47}} (1993) 1093.
\vspace*{-0.25cm}
\bibitem{Tonnis}
T. Clark, S. Love and W. ter Veldhuis,
{\it{Mod. Phys. Lett.}} {\bf{A6}} (1991) 3225.
\vspace*{-0.25cm}
\bibitem{RadCorr}
Y. Okada, M. Yamaguchi and T. Yanagida,
{\it{Prog. Theor. Phys.}} {\bf{85}} (1991) 1,
{\it{Phys. Lett.}} {\bf{B262}} (1991) 54; \newline
J. Ellis, G. Ridolfi
and  F. Zwirner,
{\it{Phys. Lett.}} {\bf{B 257}} (1991) 83; \newline
H. Haber and R. Hempfling,
{\it{Phys. Rev. Lett.}} {\bf{66}} (1991) 1815;\newline
R. Barbieri, M. Frigeni and F. Caravaglios,
{\it{Phys. Lett.}} {\bf{B258}} (1991) 167.
\vspace*{-0.25cm}
\bibitem{Higgs} M. Carena, K. Sasaki and C.E.M. Wagner,
{\it{Nucl. Phys.}} {\bf{B381}} (1992) 66.\\
P. Chankowski, S. Pokorski and J. Rosiek,
{\it{Phys. Lett.}} {\bf{B281}} (1992) 100.\\
H. Haber and R. Hempfling, {\it{UC, Santa Cruz
preprint}}, SCIPP-92-33, submitted to Phys. Rev. D.
\vspace*{-0.25cm}
\bibitem{DFS} G. Degrassi, S. Fanchiotti and A. Sirlin,
{\it{Nucl. Phys.}} {\bf{B351}} (1991) 49.
\vspace*{-0.25cm}
\bibitem{NA} R. Arnowitt and P. Nath,
{\it{Phys. Rev. Lett.}} {\bf{69}} (1992)
725; P. Nath and R. Arnowitt,
{\it{Phys. Lett.}} {\bf{B287}} (1992) 89; {\it{ibid}}
{\bf{B289}} (1992) 368.


\end{thebibliography}
\end{document}